\documentclass[12pt]{article}
\usepackage{amsmath}
\usepackage{amsfonts}

 \makeatletter
\@addtoreset{equation}{section}

\makeatother
\input{tcilatex}

\begin{document}

\bigskip \thispagestyle{empty}

\begin{center}
\null\vspace{-1cm} \hfill \\[0pt]
\vspace{1cm} \ {\Large \textbf{PSEUDO-DIFFERENTIAL OPERATORS\\[0pt]
AND \\[0pt]
INTEGRABLE MODELS\\[0pt]
}} \vspace{2cm} {\Large \textbf{M. B. Sedra}}\footnote{%
msedra@ictp.it} \\[0pt]
\textit{Laboratoire de Physique de La Mati\`ere et Rayonnement (LPMR),\\[0pt]
Univ. Ibn Tofail, Facult\'e des Sciences, K\'{e}nitra (UIT-FSK), Morocco }
\end{center}

\vspace{0.5cm} \baselineskip=18pt

The importance of the theory of pseudo-differential operators in the study
of non linear integrable systems is point out. Principally, the algebra $\Xi 
$ of nonlinear (local and nonlocal) differential operators, acting on the
ring of analytic functions $u_{s}(x, t)$, is studied. It is shown in
particular that this space splits into several classes of subalgebras $%
{\huge \Sigma }_{jr},j=0,\pm 1,r=\pm 1$ completely specified by the quantum
numbers: $s$ and $(p,q)$ describing respectively the conformal weight (or
spin) and the lowest and highest degrees. The algebra ${\huge \Sigma }_{++}$
(and its dual ${\huge \Sigma }_{--}$) of local (pure nonlocal) differential
operators is important in the sense that it gives rise to the explicit form
of the second hamiltonian structure of the KdV system and that we call also
the Gelfand-Dickey Poisson bracket. This is explicitly done in several
previous studies, see for the moment \cite{bak, GN, ss94}. Some results
concerning the KdV and Boussinesq hierarchies are derived explicitly.

\newpage \newpage

\section{Introduction}

It's well known that integrable models \cite{1}, a subject attracting much
more attention for several years both in physics and mathematics, are non
linear hamiltonian systems with an appropriately large number of conserved
quantities in involution making the system soluble.

In parallel to integrable models, another subject much more rich in
structure, deal with $2d$ conformal field theories (CFT) \cite{cft} showing
to be relevant in statistical mechanics and string theory \cite{string}. The
relation between these two classes of theories, namely the integrable models
and the CFT's would be quite interesting and useful. For instance we recall
that the KdV equation , a prototype of integrable models in two dimension
can be obtained as a $Diff S^1$ flow if we identify the Hill's operator $%
L=\partial^2+u_2$ with the space of quadratic differentials \cite{bak}. We
know also that the second hamiltonian structure of the KdV system is nothing
but the classical form of the Virasoro algebra or conformal symmetry \cite%
{GN}. In this context, we know that the KdV integrable model 
\begin{equation}
\frac{\partial}{\partial t}u(x)=u\frac{\partial}{\partial x}u+\frac{%
\partial^3 u}{\partial x^3}
\end{equation}
is a bi-Hamiltonian system in the sense that we can associate to the KdV
equation two kind of Poisson brackets called the first and the second
Hamiltonian structures. We have at equal times 
\begin{equation}
\{u(x), u(y)\}_1=\frac{\partial}{\partial x}\delta(x-y)
\end{equation}
and 
\begin{equation}
\{u(x), u(y)\}_2=(u(x)+u(y))\frac{\partial}{\partial x}\delta(x-y)+\frac{1}{2%
}\frac{\partial^3}{\partial x^3}\delta(x-y)
\end{equation}%
The bracket $\{u(x), u(y)\}_2$ is nothing but the form reproducing the
classical Virasoro algebra of the spin two algebra. On the other hand, using
(1.1) with 
\begin{equation}
H_3=\int {dx(\frac{1}{3!}u^3-\frac{1}{2}(\frac{\partial u}{\partial x})^2)}
\end{equation}
then it's easily verified that 
\begin{equation}
\frac{\partial u}{\partial t}=\{u(x), H_3\}_1=u\frac{\partial}{\partial x}u+%
\frac{\partial^3 u}{\partial x^3}
\end{equation}%
showing that the KdV equation is Hamiltonian, see \cite{dubrovin}.

In the language of $2d$ conformal field theory, the above mentioned currents 
$u_s$ are taken in general as primary $w_s$ satisfying the OPE 
\begin{equation}
T(z)w_{s}(\omega)=\frac{s}{(z-\omega)^{2}}w_{s}(\omega)+\frac{%
w^{\prime}_{s}(\omega)}{(z-\omega)},
\end{equation}
or equivalently, 
\begin{equation}
w_s=J^{s}.{\tilde w}_s
\end{equation}
under a general change of coordinate (diffeomorphism) $x \rightarrow {\tilde
x}(x)$ with $J=\frac{\partial \tilde x}{\partial x}$ is the associated
Jacobian.\newline
\newline
These $w$-symmetries \cite{w1, w2, w3} exhibit among other a non linear
structure and are not Lie algebra in the standard way as they incorporate
composite fields in their OPE. \newline
\newline
In integrable models, and as signaled previously, these higher spin
symmetries appear such that the Virasoro algebra $W_2$ defines the second
Hamiltonian structure for the KdV hierarchy, $W_3$ for the Boussinesq and $%
W_{1+\infty}$ for the KP hierarchy and so one. These correspondences are
achieved naturally in terms of pseudo-differential Lax operators. 
\begin{equation}
{\mathcal{L}}_{n}=\sum_{j\in Z}u_{n-j}\partial^{j},
\end{equation}
allowing both positive as well as nonlocal powers of the differential $%
\partial ^{j}$. The fields $u_j$ of arbitrary conformal spin $j$ did not
define a primary basis. The construction of primary fields from the $u_j$
one's is originated from the well known covariantization method of
Di-Francesco -Itzykson-Zuber (DIZ)\cite{DIZ} showing that the primary $W_j$
fields are given by adequate polynomials of $u_j$ and their k-th derivatives 
$u_{j}^{(k)}$. \newline
\newline

\section{Basics notions and convention notations}

\subsection{The algebra of currents $u_s(x,t)$}

We present in this first subsection the general setting of the basic
properties of the algebra of currents, that we usually denote as $u_s(x,t)$.
These are mathematical objects, of physical meaning, characterized by a
quantum number namely the conformal weight (spin) $s>1$. This particular
algebra is considered to be a semi-infinite dimensional space of huge
infinite tensor algebra of arbitrary integer spin fields.

The currents $u_s(x,t)$ are playing a crucial role in physics through the
conformal symmetry and its higher spin extensions called $w-$symmetry
signaled in the introduction. These extended symmetries deal with analytic
fields obeying a nonlinear closed algebra. They appear also in the study of
higher differential operators involved in the analysis of nonlinear
integrable models to be considered in the forthcoming parts of this work. We
shall start however by defining our convention notations.

The two dimensional Euclidean space $\mathbb{R}^{2}\cong \mathbb{C}$ is
parametrized by the complex coordinates $z=t+ix$ and $\bar{z}=t-ix$. As a
matter of convention, we set $z=z^{+}$ and $z=z^{-}$ so that the derivatives 
$\partial /\partial z$ and $\partial /\partial \bar{z}$ are, respectively,
represented by $\partial _{+}=\partial $ and $\partial _{-}$=$\bar{\partial}$%
. The $so(2)$ Lorentz representation fields are described by one component
tensors of the form $\psi _{k}(z,\bar{z})$ with $2k\in \mathbb{Z}$. $\mathbb{%
Z}$ is the set of relative integers. In two dimensional conformal field
theories (CFT)\cite{1}, an interesting class of fields is given by the set
of analytic fields $\phi _{k}(z)$. These are $SO(2)\cong U(1)$ tensor fields
that obey the analyticity condition $\partial _{-}\phi _{k}(z)=0$. In this
case the conformal spin $k$ coincides with the conformal dimension $\Delta$.
Note that under a $U(1)$ global transformation of parameter $\theta $, the
object $z^{\pm }$, $\partial _{\pm }$ and $\phi _{k}(z)$ transform as%
\begin{equation}
z^{\pm \prime }=e^{\mp i\theta }z^{\pm },\quad \partial _{\pm }^{\prime
}=e^{\pm i\theta }\partial _{\pm },\quad \phi _{k}^{\prime }(z)=e^{ik\theta
}\phi _{k}(z)  \label{2.1}
\end{equation}%
so that $dz\partial _{z}$ and $(dz)^{k}\phi _{k}(z)$ remain invariant. In a
pure bosonic theory, which is the purpose of the present study, only integer
values of conformal spin $k$ are involved. We denote by $\Xi ^{(0,0)}$ the
tensor algebra of analytic fields of arbitrary conformal spin. This is a
completely reducible infinite dimensional $SO(2)$ Lorentz representation
(module) that can be written as%
\begin{equation}
\Xi ^{(0,0)}=\underset{k\in \mathbb{Z}}{\oplus }\Xi _{k}^{(0,0)}  \label{2.2}
\end{equation}%
where the $\Xi _{k}^{(0,0)}$'s are one dimensional $SO(2)$ spin $k$
irreducible modules. The upper indices $(0,0)$ carried by the spaces
figuring in Eq. (\ref{2.2}) are special values of general indices $(p,q)$ to
be introduced later on. The generators of these spaces are given by the spin 
$k$ analytic fields $u_{k}(z)$. They may be viewed as analytic maps $u_{k}$
which associate to each point $z$, on the unit circle $S^{1}$, the fields $%
u_{k}(z)$. For $k\geq 2$, these $u_{k}$ fields can be thought of as the
higher spin currents involved in the construction of the higher spin
conformal currents or $w$-algebras. As an example, the following fields:%
\begin{equation}
w_{2}=u_{2}(z),\quad w_{3}=u_{3}(z)-\frac{1}{2}\partial _{z}u_{2}(z)
\label{2.3}
\end{equation}%
are the well-known spin-$2$ and spin-$3$ conserved currents of the
Zamolodchikov $w_{3}$-algebra \cite{w1, w2, w3}. As in infinite dimensional
spaces, elements $\Phi $ of the spin tensor algebra $\Xi ^{(0,0)}$ in Eq. (%
\ref{2.2}) are built from the vector basis $\{u_{k},k\in \mathbb{Z}\}$ as
follows:%
\begin{equation}
\Phi =\sum_{k\in \mathbb{Z}}c(k)u_{k}  \label{2.4}
\end{equation}%
where only a finite number of the decomposition coefficients $c(k)$ is
nonvanishing. Introducing the following scalar product $\left\langle
,\right\rangle $ in the tensor algebra $\Xi ^{(0,0)}$%
\begin{equation}
\left\langle u_{l},u_{k}\right\rangle =\delta _{k+l,1}\int
dz\,u_{1-k}(z)u_{k}(z)  \label{2.5}
\end{equation}%
it is not difficult to see that the one dimensional subspaces $\Xi
_{k}^{(0,0)}$ and $\Xi _{1-k}^{(0,0)}$ are dual to each other. As a
consequence the tensor algebra $\Xi ^{(0,0)}$ splits into two semi-infinite
tensor subalgebras $\Sigma _{+}^{(0,0)}$ and $\Sigma _{-}^{(0,0)}$,
respectively, characterized by positive and negative conformal spins as
shown here below%
\begin{equation}
\Sigma _{+}^{(0,0)}=\underset{k\succ 0}{\oplus }\Xi _{k}^{(0,0)}  \label{2.6}
\end{equation}%
\begin{equation}
\Sigma _{-}^{(0,0)}=\underset{k\succ 0}{\oplus }\Xi _{1-k}^{(0,0)}
\label{2.7}
\end{equation}%
From these equations we read in particular that $\Xi _{0}^{(0,0)}$ is the
dual of $\Xi _{1}^{(0,0)}$ and if half integers were allowed, $\Xi
_{1/2}^{(0,0)}$ would be self dual with respect to the form (\ref{2.5}).
Note that the product (\ref{2.5}) carries a conformal spin structure since
from dimensional arguments, it behaves as a conformal object of weight $%
\Delta =-1$%
\begin{equation}
\Delta \lbrack \left\langle u_{1-k},u_{k}\right\rangle ]=-1+(1-k)+k=0
\label{2.8}
\end{equation}%
Later on, we shall introduce a combined product $\left\langle \left\langle
,\right\rangle \right\rangle $ built out of Eq. (\ref{2.5}) and a pairing
product $(,)$, see Eq. (\ref{3.34}), of conformal weight $\Delta =1 $ so
that we get%
\begin{equation}
\Delta \left[ \left\langle \left\langle ,\right\rangle \right\rangle \right]
=-1+1=0  \label{2.9}
\end{equation}%
Moreover, the infinite tensor algebra $\Sigma _{+}^{(0,0)}$ of Eq. (\ref{2.6}%
) contains, in addition to the spin-l current, all the $W_{n}$ currents $%
n\geq 2$. These fields are used in the construction of higher spin local
differential operators as it will be shown later on. Analytic fields with
negative conformal spins [Eq. (\ref{2.7})] are involved in the building of
nonlocal pseudodifferential operators. Both these local and nonlocal
operators are needed in the derivation of the classical $w_{n}$-algebras
from the Gelfand-Dickey algebra of $sl(n)$.

\subsection{Introducing pseudo-operators}

Before going into more details in the next parts of this work, let's first
start by a brief account of the basic properties of the space of higher
order differential Lax operators. We have to fix that every
pseudo-differential operator is completely specified by a conformal spin $s$%
, $s\in \mathbb{Z}$, two integers $p$ and $q=p+n$, $n\geq 0$ defining the
lowest and the highest degrees respectively and finally $(1+q-p)=n+1$
analytic fields $u_{j}(z)$ see for the moment \cite{ss94}

We denote by $\mathcal{A}$ the huge algebra of all local and non local
differential operators of arbitrary conformal spins and arbitrary degrees.
One may expand $\mathcal{A}$ as 
\begin{equation}
\mathcal{A=}\underset{p\leq q}{\mathcal{\oplus }}\mathcal{A}^{(p,q)}\mathcal{%
=}\underset{p\leq q}{\mathcal{\oplus }}\underset{}{\quad ~}\underset{s\in 
\mathbb{Z}}{\mathcal{\oplus }}\mathcal{A}_{s}^{(p,q)},~p,q,s\in \mathbb{Z}
\end{equation}%
where we have denoted by $(p,q)$ the lowest and thee highest degrees
respectively and by $s$ the conformal spin. The vector space $\mathcal{A}%
^{(p,q)}$ of differential operators with given degrees $(p,q)$ but undefined
spin exhibits a Lie algebra structure with respect to the Lie bracket for $%
p\leq q\leq 1$. To see this, let us consider the set $\mathcal{A}%
_{s}^{(p,q)} $ of differential operators type 
\begin{equation}
d_{s}^{(p,q)}:=\sum_{i=p}^{q}u_{s-i}(z)\partial ^{i}
\end{equation}%
It's straightforward to check that the commutator of two operators $%
d_{s}^{(p,q)}$ is an operator of conformal spin $2s$ and degrees $(p,2q-1)$.
Since the Lie bracket $[.,.]$ acts as 
\begin{equation}
\lbrack .,.]:\mathcal{A}_{s}^{(p,q)}\times \mathcal{A}_{s}^{(p,q)}%
\longrightarrow \mathcal{A}_{2s}^{(p,2q-1)}
\end{equation}

Imposing the closure, one gets strong constraints on the spin s and the
degrees parameters (p,q) namely 
\begin{equation}
s=0\quad \text{and}\quad p\leq q\leq 1
\end{equation}

From these equations we learn in particular that the spaces $\mathcal{A}%
_{0}^{(p,q)},p\leq q\leq 1$ admit Lie algebra structure with respect to the
bracket Eq(2.13) provided that the Jacobi identity is fulfilled. This can be
ensured by showing that the Leibnitz product is associative. Indeed given
three arbitrary differential operators $%
d_{m_{1}}^{(p_{1},q_{1})},d_{m_{2}}^{(p_{2},q_{2})}$ and $%
d_{m_{3}}^{(p_{3},q_{3})}$ we find that associativity follows by help of the
identity 
\begin{equation}
\sum_{l=0}^{i}\binom{i}{l}\binom{j}{k-l}=\binom{i+j}{k},
\end{equation}%
where $\binom{i}{j}$ is the usual binomial coefficient.

\subsection{Some non linear differential equations}

\textbf{The KP equation}\newline
\begin{equation}
\partial_x (\frac{\partial u}{\partial t} -\frac{1}{4} u^{\prime\prime%
\prime}- 3uu^{\prime}) =\frac{3}{4}\frac{\partial^2 u}{\partial y^2}.
\end{equation}
\textbf{The non linear schrodinger equation} \cite{SE1, SE2}\newline
\begin{equation}
\begin{array}{lcl}
i\partial_t q & = & -q_{xx}+2k(q^{\ast}{q})q \\ 
&  &  \\ 
i\partial_t q^{\ast} & = & -q^{\ast}_{xx}-2k(q^{\ast}{q})q^{\ast},%
\end{array}
\tag{21}
\end{equation}
where $k$ is an arbitrary parameter measuring the strength of the non linear
interaction and can be set to unity through a resealing of the dynamical
variables $q$ and $q^{\ast}$.\newline
\textbf{The two bosons Hierarchy}\cite{TB}:\newline
\begin{equation}
\begin{array}{lcl}
\partial_t u & = & (2h+u^2-u_x)_x, \\ 
&  &  \\ 
\partial_t h & = & (2uh+h_x)_x%
\end{array}
\tag{21}
\end{equation}
This hierarchy is known to yields the non linear schrodinger equation once
we set $u=-\frac{q_x}{q}$ and $h=-q^{\ast}{q}$.\newline
\textbf{The Liouville equation:\newline
} 
\begin{equation}
\bar\partial\partial \phi = exp(2\phi)
\end{equation}
where $\phi$ is a Liouville scalar field. This is a conformally invariant
field theory shown to be intimately related to the KdV equation thought the
Virasoro symmetry.

\subsection{Some pseudo differential operators}

The algebra of pseudo-differential operators play fundamental role for the
construction of the Kadomtsev-Petviashvili (KP) hierarchy in the Lax
formalism. For the standard KP hierarchy, the associated pseudo-differential
operator can be regarded as an ordinary integral operator, which enjoys the
generalized Leibniz rule. We define the Lax operator of the KP hierarchy by%
\begin{equation}
\mathcal{L}_{KP}=\partial +\sum_{j=1}^{\infty }u_{j+1}\partial ^{-j},
\label{75}
\end{equation}%
where $u_{j}$'s are dependent variables of space $x$ and time variables
being introduced below. The coefficient of $\partial ^{0}$ can be set zero
without loss of generality. We assign the degree of the differential
operator $\partial $ one, standing for $\deg [\partial ]=1$, and assume that
all the terms in the Lax operator have equal degree, \textit{i.e.}, $\deg
[u_{j}]=j$. Hence the algebra wrapping this operator is given by a quotient%
\begin{equation}
\mathcal{L}_{KP}\in \mathcal{A}_{1}^{(-\infty ,1)}/\mathcal{A}_{1}^{(0,0)}
\end{equation}%
with $\mathcal{A}_{1}^{(0,0)}$ (subalgebra of algebra $\mathcal{A}%
_{1}^{(-\infty ,1)}$) the ring of the functions of spin 1. The second class
of the important pseudo differential operators is given by the Burgers
operators%
\begin{equation}
\mathcal{L}_{Burgers}=\partial _{x}+u_{1}(x,t)
\end{equation}%
which is an element of the algebra 
\begin{equation}
\mathcal{A}_{1}^{(0,1)}\equiv \mathcal{A}_{1}^{(-\infty ,1)}/\mathcal{A}%
_{1}^{(-\infty ,-1)}.
\end{equation}

Another class of the important pseudo differential operators is given by the
Korteweg-de Vries (KdV) operators%
\begin{equation}
\mathcal{L}_{KdV}=\partial _{x}^{2}+u_{2}(x,t)
\end{equation}%
the algebraic structure of this operator is given by the quotient%
\begin{equation}
\mathcal{A}_{2}^{(0,2)}/\mathcal{A}_{2}^{(1,1)}
\end{equation}%
this operator can be seen as the second reduction of the KP hierarchy, indeed%
\begin{equation}
\mathcal{L}_{KdV}=\left( \mathcal{L}_{KP}^{2}\right) _{\geq 0}=\left( 
\mathcal{L}_{KP}^{2}\right) _{+}
\end{equation}%
$\mathcal{L}_{KdV}$ is the part with only derivatives, \textit{i.e.},
non-negative power terms in $\partial $. The same formalism applies for the
Boussinesq operator which has the following form 
\begin{equation}
L_{Bous\sin esq}=\partial _{x}^{3}+u_{2}(x,t)\partial _{x}+u_{3}(x,t)
\end{equation}%
which is an element of the following algebra 
\begin{equation}
\mathcal{A}_{3}^{(0,3)}/\mathcal{A}_{3}^{(2,2)}
\end{equation}%
$\mathcal{L}_{Bous\sin esq}$ can be seen as the third is the reduction of
the KP hierarchy%
\begin{equation}
\mathcal{L}_{Bous\sin eq}=\left( \mathcal{L}_{KP}^{3}\right) _{\geq
0}=\left( \mathcal{L}_{KP}^{3}\right) _{+}
\end{equation}

\section{Pseudo-differential operator's theory \protect\cite{ss94}}

Here we describe, in a little bit more details, the basic features of
nonlinear pseudo-differential operators on the ring of analytic functions.
As quoted before, we show in particular that any such differential operator
is completely specified by a weight (spin) $m,m\in \mathbb{Z}$, two integers 
$p$ and $q=p+n,n\geq 0$ defining the lowest and highest degrees,
respectively, and finally $(1+q-p)=n+1$ analytic fields $u_{j}(z)$. We also
show that the set $\Xi $ of all nonlinear differential operators admits a
Lie algebra structure with respect to the commutator of differential
operators built out of the Leibnitz product. Moreover we find that $\Xi $
splits into $3\times 2=6$ subalgebras $\Sigma _{j+}$ and $\Sigma _{j-}$ $,$ $%
j=0,\pm 1$ related to each other by two types of conjugations, namely, the
spin and degree conjugations. The algebras $\Sigma _{++}$ and $\Sigma _{--}$
are of particular interest in this study as they are used in the
construction of the Hamiltonian structure of nonlinear integrable models.

To that purpose we shall proceed as follows: First we introduce the space of
differential operators of fixed spin $m$ and fixed degrees $(p,q)$. This
space is referred hereafter to as $\Xi _{m}^{(p,q)}$. Then we consider the
set $\Xi ^{(p,q)}$ of nonlinear operators of fixed degrees $(p,q)$ but
arbitrary spin. Finally we build the desired space.

\subsection{The $\Xi _{m}^{(p,q)}$ space}

To begin, we remark that $\Xi _{m}^{(p,q)}$ is the space of differential
operators whose elements $d_{m}^{(p,q)}(u)$ are the generalization of the
well-known scalar Lax operator involved in the analysis of the so-called KdV
hierarchies and in Toda theories \cite{tod1, tod2, tod3}. The simplest
example is given by the Hill operator%
\begin{equation}
L=\partial ^{2}+u(z)  \label{3.1}
\end{equation}%
which plays an important role in the study of the Liouville theory and in
the KdV equation. A natural generalization of the above relation is given by%
\begin{equation}
d_{m}^{(p,q)}(u)=\sum_{i=p}^{q}u_{m-i}(z)\partial ^{i}  \label{3.2}
\end{equation}%
where the $u_{m-i}(z)$'s are analytic fields of spin $(m-i)$, $p$ and $q$,
with $p\geq q$, are integers that we suppose are positive for the moment. We
shall refer hereafter to $p$ as the lowest degree of $d_{m}^{(p,q)}(u)$ and $%
q$ as the highest one. We combine these two features of Eq. (\ref{3.2}) by
setting%
\begin{equation}
Deg(d_{m}^{(p,q)}(u))=(p,q)  \label{3.3}
\end{equation}%
m is the conformal spin of the $(1+q-p)$ monomes of the right hand side
(rhs) of Eq. (\ref{3.2}) and then of $d_{m}^{(p,q)}(u)$ itself. As for the
above relation, we set%
\begin{equation}
\Delta (d_{m}^{(p,q)}(u))=m  \label{3.4}
\end{equation}%
Putting $m=2,p=0$, and $q=2$ together with the special choices $u_{0}(z)=1$
and $u_{1}(z)=0$ in Eq. (\ref{3.2}), we recover Eq. (\ref{3.1}) as a
particular object. Moreover, Eq. (\ref{3.2}) which is well defined for $%
q\geq p\geq 0$ may be extended to negative integers by introducing
pseudodifferential operators of the type $\partial ^{-k}$, $k\geq 0$, whose
action on the fields $u_{s}(z)$ is defined as%
\begin{equation}
\partial ^{-k}u_{s}(z)=\sum_{l=0}^{\infty
}(-1)^{l}C_{k+l-1}^{l}u_{s}^{(l)}(z)\partial ^{-k-l}  \label{3.5}
\end{equation}%
where $u_{s}^{(l)}(z)$ is the $l-$th derivative of $u_{s}(z)$. As can be
checked by using the Leibnitz rule, Eq. (\ref{3.5}) obeys the expected
property%
\begin{equation}
\partial ^{k}\partial ^{-k}u_{s}(z)=u_{s}(z)  \label{3.6}
\end{equation}%
A natural representation basis of nonlinear pseudodifferential operators of
spin $m$ and negative degrees $(p,q)$ is given by%
\begin{equation}
\delta _{m}^{(p,q)}(u)=\sum_{i=p}^{q}u_{m-i}(z)\partial ^{i}  \label{3.7}
\end{equation}%
This configuration, which is a direct extension of Eq. (\ref{3.2}), is
useful in the study of the algebraic structure of the spaces $\Xi
_{m}^{(p,q)}$ and $\Xi ^{(p,q)}$. Note by the way that we can use another
representation of pseudodifferential operators, namely, the Volterra
representation. The latter is convenient in the derivation of the second
Hamiltonian structure of higher conformal spin integrable theories. Note
also that Eq. ({3.5}) is a special pseudodifferential operator of the type
of Eq. (\ref{3.7}) with $m=s-k,p=-\infty $ and $q=-k$.

Using Eqs. (\ref{3.2}) and (\ref{3.7}), one sees that operators with
negative lowest degrees $p$ and positive highest degrees $q$ denoted by $%
D_{m}^{(p,q)}[u]$ split as%
\begin{equation}
D_{m}^{(p,q)}\left[ u\right] =\delta _{m}^{(p,q)}(u)+d_{m}^{(p,q)}(u)
\label{3.8}
\end{equation}%
More generally we have%
\begin{equation}
D_{m}^{(p,q)}\left[ u\right] =D_{m}^{(p,k)}(u)+D_{m}^{(k+1,q)}(u)
\label{3.9}
\end{equation}%
for any integers $p\leq k\leq q$. As a consequence, one finds that the
operation (\ref{3.3}) obeys the rule%
\begin{equation}
(p,q)=(p,k)+(k+1,q)  \label{3.10}
\end{equation}%
for any three integers such that $p\leq k\,<q$. Now let $\Xi _{m}^{(p,q)}$; $%
m,p,$ and $q$ integers with $q\geq p$, be the set of spin $m$ differential
operators of degrees $(p,q)$. With respect to the usual addition and
multiplication by $C$ numbers, $\Xi _{m}^{(p,q)}$ behaves as $(1+q-p)$
dimensional space generated by the vector basis%
\begin{equation*}
\left\{ D_{m}^{(p,q)}\left[ u\right] =\sum_{i=p}^{q}u_{m-i}\partial
^{i};\,\,\,\,p\leq i\leq q\right\}
\end{equation*}%
Thus the space decomposition of $\Xi _{m}^{(p,q)}$ reads as%
\begin{equation}
\Xi _{m}^{\left( p,q\right) }=\underset{i=p}{\overset{q}{\oplus }}\Xi
_{m}^{(i,i)}  \label{3.11}
\end{equation}%
where the $\Xi _{m}^{(i,i)}$'s are one dimensional spaces given by%
\begin{equation}
\Xi _{m}^{(i,i)}=\Xi _{m-i}^{(0,0)}\otimes \partial ^{i}  \label{3.12}
\end{equation}%
Setting $i=0$, one discovers the space $\Xi _{m}^{(0,0)}$ defined
previously. Remark that the number of independent fields $u_{j}(z)$ involved
in Eqs. (\ref{3.2}) and (\ref{3.7})-(\ref{3.9}) is equal to the dimension of 
$\Xi _{m}^{(p,q)}$. In the next subsection we shall show that among all the
spaces $\Xi _{m}^{(p,q)}$, $m,p$ and $q$ arbitrary integers, only the sets $%
\Xi _{0}^{(p,q)}$ with $p<q<1$ admit a Lie algebra structure with respect to
the bracket $[D_{1},D_{2}]=D_{1}\circ D_{2}-D_{2}\circ D_{1}$ constructed
out of the Leibnitz product $\circ $. Because of the relations (\ref{3.9})-(%
\ref{3.10}) these spaces obey%
\begin{equation}
\Xi _{0}^{(p,k)}\subset \Xi _{0}^{(p,q)},\quad \Xi _{0}^{(k+1,q)}\subset \Xi
_{0}^{(p,q)}  \label{3.13}
\end{equation}%
and are then subalgebras of the maximal Lie algebra $\Xi _{0}^{(-\infty ,1)}$%
\ of Lorentz scalar differential operators of highest degree less than 2.

\subsection{Explicit formulas}

Performing computations, we find the following Leibnitz rules: 
\begin{equation}
\begin{array}{lcl}
\partial u & = & u\partial+ u^{\prime }, \\ 
&  &  \\ 
\partial^{2}u & = & u \partial^{2}+2 u^{\prime }\partial+ u^{\prime \prime },
\\ 
&  &  \\ 
\partial^{3}u & = & u \partial^{3}+3 u^{\prime }\partial^{2}+3 u^{\prime
\prime }\partial+ u^{\prime \prime \prime }, \\ 
&  & 
\end{array}
\tag{21}
\end{equation}
and 
\begin{equation}
\begin{array}{lcl}
\partial^{-1}u & = & u \partial^{-1}- u^{\prime }\partial^{-2}+ u^{\prime
\prime }\partial^{-3}-u^{\prime \prime \prime }\partial^{-4}+..., \\ 
&  &  \\ 
\partial^{-2}u & = & u \partial^{-2}-2u^{\prime }\partial^{-3}+3u^{\prime
\prime }\partial^{-4}-4(\frac{1}{2})^{3}u^{\prime \prime \prime
}\partial^{-5}+..., \\ 
&  &  \\ 
\partial^{-3} u & = & u\partial^{-3}-3\frac{1}{2} u^{\prime }\partial^{-4}+6(%
\frac{1}{2})^{2}u^{\prime \prime }\partial^{-5}-10(\frac{1}{2})^{3}u^{\prime
\prime \prime }\partial^{-6}+.... \\ 
&  & 
\end{array}
\tag{22}
\end{equation}
The general expressions are given, for positive value of $n$, by the
Leibnitz rules 
\begin{equation}
\begin{array}{lcl}
\partial^{n} u & = & \sum_{i=0}^{n}c_{n}^{i}u^{(i)}\partial^{n-i}, \\ 
&  &  \\ 
\partial^{-n}u & = & \sum_{i=0}^{\infty
}(-1)^{i}c_{n+i-1}^{i}u^{(i)}\partial^{-n-i}.%
\end{array}
\tag{23}
\end{equation}

\subsection{The $\Xi _{0}^{(p,q)}$ Lie algebra}

We start by considering differential operators of the type of Eq.(\ref{3.2}%
). The Leibnitz product 0 acts on $\Xi _{m}^{(p,q)}$; $q\geq p\geq 0$ as%
\begin{equation}
\circ :\Xi _{m}^{(p,q)}\times \Xi _{m}^{(p,q)}\longrightarrow \Xi
_{2m}^{(p,2q)}  \label{3.14}
\end{equation}%
or equivalently%
\begin{equation}
d_{m}^{(p,q)}(u)\times d_{m}^{(p,q)}(v)=\sum_{k=p}^{2q}w_{2m-k}\partial
^{k}=d_{2m}^{(p,2q)}(w)  \label{3.15}
\end{equation}%
where the composite analytic fields $w_{2m-k}$ are given by%
\begin{equation}
w_{2m-k}(z)=\sum_{i=p}^{q}\sum_{j=p}^{q}C_{i}^{k-j}u_{m-i}v_{m-j}^{\left(
i+j-k\right) },\,\,\,C_{r}^{s}=\frac{r!}{s!\left( r-s!\right) },\,\,0\leq
s\leq r.  \label{3.16}
\end{equation}%
Similar relations may be written down for the commutator of two differential
operators. We have%
\begin{equation}
\left[ ,\right] :\Xi _{m}^{(p,q)}\times \Xi _{m}^{(p,q)}\longrightarrow \Xi
_{2m}^{(p,2q-1)}  \label{3.17}
\end{equation}%
showing that in general the space $\Xi _{m}^{(p,q)}$ is not closed under the
action of the bracket $\left[ ,\right] $. Imposing the closure, one gets
strong constraints on the integers $m,p,$ and $q,$ namely,%
\begin{equation}
m=0,\quad 0\leq p\leq q\leq 1  \label{3.18}
\end{equation}%
The spaces $\Xi _{0}^{(p,q)}$ satisfying the above constraint equations then
exhibit a Lie algebra structure provided that the Jacobi identity is
fulfilled. This can be ensured by showing that the Leibnitz product is
associative. Indeed given three arbitrary differential operators%
\begin{equation}
D_{1}=d_{m_{1}}^{\left( p_{1},q_{1}\right) },\quad D_{2}=d_{m_{2}}^{\left(
p_{2},q_{2}\right) },\quad D_{3}=d_{m_{3}}^{\left( p_{3},q_{3}\right) }
\label{3.19}
\end{equation}%
we find that associativity%
\begin{equation}
D_{1}\circ \left( D_{2}\circ D_{3}\right) =\left( D_{1}\circ D_{2}\right)
\circ D_{3}  \label{3.20}
\end{equation}%
follows by with help of the identity%
\begin{equation}
\sum_{l=0}^{i}C_{i}^{l}C_{j}^{k-l}=C_{i+j}^{k}  \label{3.21}
\end{equation}%
Note that the constraint relations (\ref{3.18}) give rise to three Lie
algebras $\Xi _{0}^{(0,0)}$, $\Xi _{0}^{(1,1)}$, and $\Xi _{0}^{(0,1)}$
obeying the obvious space decomposition%
\begin{equation}
\Xi _{0}^{(0,0)}\oplus \Xi _{0}^{(1,1)}=\Xi _{0}^{(0,1)}  \label{3.22}
\end{equation}%
Here, one recognizes $\Xi _{0}^{(1,1)}$\ as just the Lie algebra of vector
fields on the circle, namely, $Diff(S^{1})$. $\Xi _{0}^{(0,0)}$ is the one
dimensional trivial $\Xi _{0}^{(0,1)}$\ ideal as that in Eq. (\ref{3.22}).
The above analysis can be extended to the pseudodifferential operators $%
\delta _{m}^{(p,q)}(u)$ ; $p\leq q<0$. Using analogous calculations, we find
that the spaces $\Xi _{m}^{(p,q)}$ are closed with respect to the commutator 
$[,]$ provided that%
\begin{equation}
m=0,\quad \left( p+1\right) /2\leq q\leq -1  \label{3.23}
\end{equation}%
Here also associativity ensures that the vector spaces $\Xi _{0}^{(p,q)}$\
satisfying Eqs. (\ref{3.23}) admit Lie algebra structures. These spaces as
well as $\Xi _{0}^{(0,1)}$ are in fact subalgebras of the huge Lie algebra $%
\Xi _{0}^{(-\infty ,1)}$. The latter has the remarkable space decomposition%
\begin{equation}
\Xi _{0}^{(-\infty ,1)}=\Xi _{0}^{(-\infty ,-1)}\oplus \Xi _{0}^{(0,1)}
\label{3.24}
\end{equation}%
where $\Xi _{0}^{(-\infty ,1)}$\ is the Lie algebra of Lorentz scalar pure
pseudodifferential operators of higher degree $q=-1$. Note finally that for
a given $k\leq 1$, we have%
\begin{equation}
\Xi _{0}^{(-\infty ,k-1)}\subset \Xi _{0}^{(-\infty ,k)}  \label{3.25}
\end{equation}%
and by Eq. (\ref{3.17})%
\begin{equation}
\left[ \Xi _{0}^{(-\infty ,l)},\Xi _{0}^{(-\infty ,k)}\right] \subset \Xi
_{0}^{(-\infty ,k+l-1)}  \label{3.26}
\end{equation}%
which in turn shows that all $\Xi _{0}^{(-\infty ,k-n)},n>0$ are ideals of $%
\Xi _{0}^{(-\infty ,k)}$. Thus the spaces $\Xi _{0}^{(p,q)}$, $p\leq q<1$,
may be viewed as coset algebras%
\begin{equation}
\Xi _{0}^{(p,q)}=\Xi _{0}^{(-\infty ,q)}/\Xi _{0}^{(-\infty ,p-1)}
\label{3.27}
\end{equation}%
Note that the space of Lorentz scalar differential operators $\Xi
_{0}^{(p,q)}$ emerges as a special set of the algebra of all nonlinear
differential operators.

\subsection{The $\Xi ^{(p,q)}$ algebra $(p\leq q\leq 1)$}

So far we have seen that the closure of the commutator of higher
differential operators imposes constraints on the conformal spin and on the
degrees, Eqs. (\ref{3.18}) and (\ref{3.23}). The restriction on the spin can
be overcome by using the spin tensor algebra $\Xi ^{(0,0)}$ given by Eq.(\ref%
{2.2}) instead of $\Xi _{0}^{(0,0)}$. As a consequence, we get a larger set
of differential operators than $\Xi _{m}^{(p,q)}$. This space to which we
refer hereafter to as $\Xi ^{(p,q)}$, $q\geq p$, is constructed as follows: 
\begin{equation}
\Xi ^{(p,q)}=\underset{m\in \mathbb{Z}}{\oplus }\Xi _{m}^{(p,q)},\quad
p,q\in \mathbb{Z}  \label{3.28}
\end{equation}%
The elements $D^{(p,q)}$ of this infinite dimensional space are differential
operators with fixed degrees $(p,q)$ but arbitrary spin. They read as%
\begin{equation}
D^{\left( p,q\right) }=\sum_{m\in \mathbb{Z}}C\left( m\right) D_{m}^{\left(
p,q\right) },  \label{3.29}
\end{equation}%
where on1y a finite number of the $c(m)$'s are not vanishing. Setting $p=q=0$%
, we get the tensor algebra $\Xi _{0}^{(0,0)}$. As for the spaces $\Xi
_{m}^{(p,q)}$, the set $\Xi ^{(p,q)}$ given by Eq. (\ref{3.28}) exhibits a
Lie algebra structure with respect to the bracket $[,]$ for p$\leq q\leq 1$.
Thus analogous relations to Eqs. (\ref{3.35})-(\ref{3.27}) can be written
down for $\Xi ^{(p,q)}$. In particular $\Xi ^{(p,q)},p\leq q\leq 1,$ may be
viewed as a coset algebra of $\Xi ^{(-\infty ,q)}$ by $\Xi ^{(-\infty ,p-1)}$%
, i.e.,%
\begin{equation}
\Xi ^{(p,q)}=\Xi ^{(-\infty ,q)}/\Xi ^{(-\infty ,p-1)}.  \label{3.30}
\end{equation}

\subsection{The algebra of differential operators $\Xi $}

This is the algebra of differential operators of arbitrary spins and
arbitrary degrees. It is obtained from Eq. (\ref{3.28}) by summing over all
the allowed degrees of the spaces $\Xi ^{(p,q)}$. In some sense, it is the
degree tensor algebra of $\Xi ^{(p,q)}$%
\begin{equation}
\Xi =\underset{p\leq q}{\oplus }\Xi ^{(p,q)}  \label{3.31}
\end{equation}%
or equivalently%
\begin{equation}
\Xi =\underset{p\in \mathbb{Z}}{\oplus }\left[ \underset{n\in \mathbb{N}}{%
\oplus }\Xi ^{(p,p+n)}\right]  \label{3.32}
\end{equation}%
Note that this infinite dimensional space is closed under the Leibnitz
commutator without any constraint. Note also that $\Xi $ is just the
combined spin and degree tensor algebra since we have%
\begin{equation}
\Xi =\underset{p\in \mathbb{Z}}{\oplus }\left[ \underset{n\in \mathbb{N}}{%
\oplus }\left( \underset{m\in \mathbb{Z}}{\oplus }\Xi _{m}^{(p,p+n)}\right) %
\right]  \label{3.33}
\end{equation}%
A remarkable property of $\Xi $ is that it splits into six infinite
subalgebras $\Sigma _{q+}$ and $\Sigma _{q-}$, $q=0,\pm 1,$ related to each
others by conjugation of the spin and degrees. Indeed given two integers $%
q\geq p$, it is not difficult to see that the spaces $\Xi ^{(p,q)}$ and $\Xi
^{(-1-q,-1-p)}$ are dual with respect to the pairing product $(,)$ defined as%
\begin{equation}
\left( D^{(r,s)},D^{\left( p,q\right) }\right) =\delta _{1+r+q,0}\delta
_{1+s+p,0}res\left[ D^{\left( r,s\right) }\circ D^{\left( p,q\right) }\right]
,  \label{3.34}
\end{equation}%
where the residue operation (res) is given by 
\begin{equation}
res\left[ \partial ^{i}\right] :=\delta _{i+1,0}.  \label{3.35}
\end{equation}%
As already shown in Eq. (\ref{2.9}), note that the operation $res$ carries a
conformal weight $\Delta =1$ and then the residue of any operator $%
D_{m}^{(p,q)}$ is%
\begin{equation}
res\left( \sum_{i=p}^{q}u_{m-i}\left( z\right) \partial ^{i}\right)
=u_{m+1}\left( z\right)  \label{3.36}
\end{equation}%
if $p\leq -1$ and $q\geq -1$ and zero elsewhere. We have, for instance,%
\begin{equation}
\begin{array}{c}
\Delta \left[ res\left( \partial ^{-1}\right) \right] =1-1=0 \\ 
\Delta \left[ res\left( u_{m+1}\partial ^{-1}\right) \right] =1+\left(
m+1\right) -1=m+1%
\end{array}
\label{3.37}
\end{equation}%
Then using Eqs. (\ref{3.34})-(\ref{3.37}), we can decompose $\Xi $ as%
\begin{equation}
\Xi =\Xi _{-}\oplus \Xi _{+}  \label{3.38}
\end{equation}%
with%
\begin{equation}
\Xi _{+}=\underset{p\geq 0}{\oplus }\left[ \underset{r\geq 0}{\oplus }\Xi
^{\left( p,p+r\right) }\right] ,  \label{3.39}
\end{equation}%
\begin{equation}
\Xi _{-}=\underset{p\geq 0}{\oplus }\left[ \underset{r\geq 0}{\oplus }\Xi
^{\left( -1-p-r,-1-p\right) }\right] .  \label{3.40}
\end{equation}%
The + and - down stairs indices carried by $\Xi _{+}$ and $\Xi _{-}$ refer
to the positive and negative degrees, respectively. Knowing that the $\Xi
^{(p,p+r)}$ spaces can also be decomposed as in Eqs. (\ref{2.6}) and (\ref%
{2.7}), with a slight modification due to Eqs. (\ref{2.8}) (\ref{2.9}), (\ref%
{3.37}), and (\ref{3.38})%
\begin{equation}
\Xi ^{(p,p+r)}=\Sigma _{-}^{(p,p+r)}\oplus \Sigma _{0}^{(p,p+r)}\oplus
\Sigma _{+}^{(p,p+r)},  \label{3.41}
\end{equation}%
where $\Sigma _{-}^{(p,p+r)}$ and $\Sigma _{+}^{(p,p+r)}$ denote the spaces
of differential operators of negative and positive definite spin. They reads
as%
\begin{equation}
\Sigma _{-}^{(p,p+r)}=\underset{m\succ 0}{\oplus }\Xi _{-m}^{\left(
p,p+r\right) },  \label{3.42}
\end{equation}%
\begin{equation}
\Sigma _{0}^{(p,p+r)}=\Xi _{0}^{\left( p,p+r\right) },  \label{3.43}
\end{equation}%
\begin{equation}
\Sigma _{+}^{(p,p+r)}=\underset{m\succ 0}{\oplus }\Xi _{m}^{\left(
p,p+r\right) },  \label{3.44}
\end{equation}%
$\Sigma _{0}^{(p,p+r)}$ is the space of Lorentz scalar differential
operators. Then plugging in Eqs. (\ref{3.38})-(\ref{3.40}), we find that z
decomposes into $3\times 2=6$ subalgebras as%
\begin{equation}
\Xi =\underset{q=0,\pm 1}{\oplus }\left[ \Sigma _{q+}\oplus \Sigma _{q-}%
\right] ,  \label{3.45}
\end{equation}%
where%
\begin{equation}
\Sigma _{q+}=\underset{p\geq 0}{\oplus }\left[ \underset{r\geq 0}{\oplus }%
\Sigma _{q}^{(p,p+r)}\right] ,\quad q=0,\pm 1  \label{3.46}
\end{equation}%
\begin{equation}
\Sigma _{q-}=\underset{p\geq 0}{\oplus }\left[ \underset{r\geq 0}{\oplus }%
\Sigma _{q}^{(-1-p-r,-1-p)}\right] .  \label{3.47}
\end{equation}%
Introducing the combined scalar product $\left\langle \left\langle
,\right\rangle \right\rangle $ built out of the product (\ref{2.5}) and the
pairing (\ref{3.34}), namely,%
\begin{equation}
\left\langle \left\langle D_{m}^{\left( r,s\right) },D_{m}^{\left(
p,q\right) }\right\rangle \right\rangle =\delta _{n+m,0}\delta
_{1+r+q,0}\delta _{1+s+p,0}\int dz\,\,\,res\left[ D_{m}^{\left( r,s\right)
}\circ D_{-m}^{\left( -1-s,-1-r\right) }\right]  \label{3.48}
\end{equation}%
one sees that $\Sigma _{++}$, $\Sigma _{0+}$ and $\Sigma _{-+}$ are the
duals of $\Sigma _{--}$, $\Sigma _{0-}$ and $\Sigma _{+-}$, respectively.
Note that $\Sigma _{0-}$ is just the algebra of Lorentz scalar
pseudo-operators of higher degree ($-1$) considered later on.

We conclude this section by making two comments: First we remark that $%
\Sigma _{++}$ is the space of local differential operators of positive
definite spins and positive degrees. $\Sigma _{--}$ however, is the Lie
algebra of nonlocal operators of negative definite spins and negative
degrees. It is these two subalgebras that will be considered in the
remainder of this study. The spaces $\Sigma _{0+}$ and $\Sigma _{0-}$ are
very special subalgebras and will be analyzed in a future article.

The second comment we want to make is to note that once we know that local
and nonlocal differential operators are completely specified by the spin and
the degrees, one may reverse the previous discussion. Start with $\Xi $, the
algebra of all possible nonlinear operators and decompose it with respect to
subspaces with definite spin and definite degrees as%
\begin{equation}
\Xi =\underset{p\in \mathbb{Z}}{\oplus }\,\,\,\underset{k\in \mathbb{Z}}{%
\oplus }\,\,\,\underset{m\in \mathbb{Z}}{\oplus }\,\Xi _{m}^{(p,p+k)}.
\label{3.49}
\end{equation}%
So that the basic objects in $\Xi $ are $\Xi _{m}^{(p,p+k)}$. All the sets
introduced above appear here as special subspaces. Their conformal as well
as the degree properties are summarized in the following table: 
\begin{equation}
\begin{tabular}{|c|c|c|}
\hline
& Conformal weight $\Delta$ & Degrees: $Deg$ \\ \hline
$\Xi _{m}^{\left( p,q\right)}$ & $m$ & $(p,q)$ \\ \hline
$\Xi^{\left( p,q\right)}$ & $indefinite$ & $(p,q)$ \\ \hline
$\left\langle ,\right\rangle _{\Xi _{m}^{\left( 0,0\right) }}$ & $-1$ & $%
(0,0)$ \\ \hline
$res$ & $1$ & $(0,0)$ \\ \hline
$\ll ,\gg$ & $0$ & $(0,0)$ \\ \hline
\end{tabular}%
\end{equation}

\section{$sl_n$ KdV-hierarchy}

The aim of this section is to present some results related to the KdV
hierarchy. Using our convention notations and the analysis that we developed
previously, we will perform hard algebraic computations and derive the KdV
hierarchy.

These computations are very hard and difficult to realize in the general
case. We will simplify this study by limiting our computations to the
leading orders of the hierarchy namely the $sl_2$-KdV and $sl_3$-Boussinesq
integrable hierarchies.

Our contribution to this study consists in extending known results by
increasing the order of computations a fact which leads us to discover more
important properties as we will explicitly show. As an original result, we
will build the deformed $sl_3$-Boussinesq hierarchy and derive the
associated flows. Some other important results are also presented.

\subsection{$sl_2$-KdV hierarchy}

Let's consider the $sl_2$ Lax operator 
\begin{equation}
{\mathcal{L}}_2= \partial^2 + u_2
\end{equation}
whose 2th root is given by 
\begin{equation}
\begin{array}{lcl}
{\mathcal{L}}^{\frac{1}{2}} & = & \Sigma_{i=-1} b_{i+1} \partial^{-i} \\ 
&  &  \\ 
& = & {\Sigma}_{i=-1} a_{i+1}\partial^{-i}%
\end{array}%
\end{equation}
This 2th root of ${\mathcal{L}}_2$ is an object of conformal spin $[{%
\mathcal{L}}^{\frac{1}{2}}]=1$ that plays a central role in the derivation
of the Lax evolutions equations.

Performing lengthy but straightforward calculations we compute the
coefficients $b_{i+1}$ of ${\mathcal{L}}^{\frac{1}{2}}$ up to $i=7$ given by 
\begin{equation}
\begin{array}{lcl}
b_{0} & = & 1 \\ 
&  &  \\ 
b_{1} & = & 0 \\ 
&  &  \\ 
b_{2} & = & \frac{1}{2}u \\ 
&  &  \\ 
b_{3} & = & -\frac{1}{4}u^{^{\prime }} \\ 
&  &  \\ 
b_{4} & = & -\frac{1}{8}u^{2}+\frac{1}{2}(\frac{1}{2})^{2}u^{^{\prime \prime
}} \\ 
&  &  \\ 
b_{5} & = & -\frac{1}{2}(\frac{1}{2})^{3}u^{^{\prime \prime \prime }}+\frac{3%
}{8}uu^{^{\prime }} \\ 
&  &  \\ 
b_{6} & = & \frac{1}{16}u^{3}-\frac{7}{4}(\frac{1}{2})^{2}uu^{^{\prime
\prime }}-\frac{11}{8}(\frac{1}{2})^{2}(u^{^{\prime }})^{2}+\frac{1}{2}(%
\frac{1}{2})^{4}u^{^{\prime \prime \prime \prime }} \\ 
&  &  \\ 
b_{7} & = & -\frac{15}{32}u^{2}u^{\prime }+(\frac{1}{2})^{3}(\frac{15}{2}%
u^{\prime \prime }u^{\prime }+\frac{15}{4}uu^{\prime \prime \prime \prime })-%
\frac{1}{2}(\frac{1}{2})^{5}u^{(5)} \\ 
&  &  \\ 
b_{8} & = & -\frac{5}{128}u^{4}+(\frac{1}{2})^{2}(\frac{55}{16}u^{\prime
\prime }u^{2}+\frac{85}{16}u{u^{\prime }}^{2})-(\frac{1}{2})^{4}(\frac{31}{4}%
uu^{\prime \prime \prime \prime }+\frac{91}{8}u^{\prime \prime 2}+\frac{37}{2%
}u^{\prime }u^{\prime \prime \prime })+\frac{1}{2}(\frac{1}{2})^{6}u^{(6)}%
\end{array}%
\end{equation}%
and 
\begin{equation}
\begin{array}{lcl}
b_{9} & = & \frac{35}{64}u^{3}u^{\prime }-\frac{175}{4}(\frac{1}{2}%
)^{3}\left( uu^{\prime }u^{\prime \prime }+\frac{1}{4}u^{\prime }{}^{3}+%
\frac{1}{4}u^{2}u^{\prime \prime \prime }\right) +\frac{7}{4}(\frac{1}{2}%
)^{5}\left( 9uu^{(5)}+25u^{(4)}u^{\prime }+35u^{\prime \prime \prime
}u^{\prime \prime }\right)  \\ 
&  &  \\ 
&  & -\frac{1}{2}(\frac{1}{2})^{7}u^{(7)} \\ 
&  &  \\ 
b_{10} & = & {\frac{7}{256}}u^{5}-\frac{35}{32}(\frac{1}{2})^{2}\left( \frac{%
23}{2}u^{2}u^{^{\prime }2}+5u^{3}u^{\prime \prime }\right) +\frac{7}{4}(%
\frac{1}{2})^{4}\left( \frac{73}{4}u^{2}u^{(4)}+\frac{227}{4}uu^{^{\prime
\prime }2}+\frac{337}{4}u^{\prime \prime }u^{^{\prime }2}+89uu^{\prime
}u^{\prime \prime \prime }\right)  \\ 
&  &  \\ 
&  & -\frac{3}{4}(\frac{1}{2})^{6}\left( \frac{631}{3}u^{^{\prime \prime
}}u^{(4)}+233u{u^{\prime \prime \prime }}^{2}+135u^{\prime }u^{(5)}\right) +%
\frac{1}{2}(\frac{1}{2})^{8}u^{(8)} \\ 
&  &  \\ 
&  & 
\end{array}%
\end{equation}%
These results are obtained by using the identification ${\mathcal{L}}_{2}={%
\mathcal{L}}^{\frac{1}{2}}\times {\mathcal{L}}^{\frac{1}{2}}$. Note that by
virtue of Eq.(4.2), the coefficients $a_{i+1}$ are shown to be functions of $%
b_{i+1}$ and their derivatives in the following way 
\begin{equation}
a_{i+1}=\sum_{s=0}^{i-1}(\frac{1}{2})^{s}c_{i-1}^{s}b_{i+1-s}^{(s)},
\end{equation}%
We obtain the results: 
\begin{equation}
\begin{array}{lcl}
a_{0} & = & 1 \\ 
&  &  \\ 
a_{2} & = & \frac{1}{2}u \\ 
&  &  \\ 
a_{4} & = & -\frac{1}{8}u^{2} \\ 
&  &  \\ 
a_{6} & = & \frac{1}{16}u^{3}+\frac{1}{8}(\frac{1}{2})^{2}(u^{\prime
}{}^{2}-2uu^{^{\prime \prime }}) \\ 
&  &  \\ 
a_{8} & = & -{\frac{5}{128}}u^{4}+{\frac{5}{8}}(\frac{1}{2})^{2}\left(
u^{2}u^{\prime \prime }-\frac{1}{2}u^{^{\prime }2}u\right) +{\frac{1}{4}}(%
\frac{1}{2})^{4}\left( u^{\prime \prime \prime }u^{\prime }-uu^{(4)}-\frac{1%
}{2}u^{^{\prime \prime }2}\right)  \\ 
&  &  \\ 
a_{10} & = & {\frac{7}{256}}u^{5}+{\frac{35}{64}}(\frac{1}{2})^{2}\left( 
\frac{1}{2}u^{2}u^{^{\prime }2}-u^{3}u^{\prime \prime }\right) +{\frac{7}{4}}%
(\frac{1}{2})^{4}\left( \frac{3}{4}u^{(4)}u^{2}+\frac{7}{4}u^{^{\prime
\prime }2}u-\frac{3}{4}u^{\prime }{}^{2}u^{\prime \prime }-uu^{\prime
}u^{\prime \prime \prime }\right)  \\ 
&  &  \\ 
& + & {\frac{1}{4}}(\frac{1}{2})^{6}\left( u^{\prime }u^{(5)}+\frac{1}{2}%
u^{\prime \prime \prime }{}^{2}-uu^{6}\right)  \\ 
&  &  \\ 
a_{12} & = & -{\frac{21}{1024}}u^{6}+{\frac{105}{64}}(\frac{1}{2})^{2}\left(
u^{4}u^{\prime \prime }-\frac{1}{2}u^{3}u^{\prime }{}^{2}\right)  \\ 
&  &  \\ 
& + & {\frac{1}{16}}(\frac{1}{2})^{4}\left( 147uu^{\prime \prime }u^{\prime
}{}^{2}+\frac{189}{2}u^{2}u^{\prime }u^{\prime \prime \prime }-\frac{1029}{4}%
u^{2}u^{^{\prime \prime }2}-63u^{3}u^{(4)}-\frac{105}{8}u^{\prime
}{}^{4}\right)  \\ 
&  &  \\ 
& + & {\frac{1}{4}}(\frac{1}{2})^{6}\left( 16u^{^{\prime \prime
}3}+9u^{2}u^{(6)}-27u^{\prime }u^{\prime \prime }u^{\prime \prime \prime }-%
\frac{45}{2}u^{\prime }{}^{2}u^{(4)}-\frac{69}{4}u^{^{\prime \prime \prime
}2}u+\frac{153}{2}uu^{\prime \prime }u^{(4)}-\frac{27}{2}uu^{\prime
}u^{(5)}\right)  \\ 
&  &  \\ 
& + & {\frac{1}{4}}(\frac{1}{2})^{8}\left( u^{\prime }u^{(7)}+u^{\prime
\prime \prime }u^{(5)}-u^{\prime \prime }u^{(6)}-uu^{(8)}-\frac{1}{2}{u^{(4)}%
}^{2}\right)  \\ 
&  & \vdots 
\end{array}%
\end{equation}%
with 
\begin{equation}
a_{2k+1}=\sum_{s=0}^{2k-1}{(\frac{1}{2})^{s}}c_{2k-1}^{s}b_{2k+1-s}^{(s)}=0,%
\hspace{1cm}k=0,1,2,3,...
\end{equation}%
Now having derived the explicit expression of ${\mathcal{L}}^{\frac{1}{2}}$,
we are now in position to write the explicit forms of the set of $sl_{n}$
KdV hierarchy. These equations defined as 
\begin{equation}
\frac{\partial {\mathcal{L}}}{\partial t_{k}}=\{({\mathcal{L}}^{\frac{k}{2}%
})_{+},{\mathcal{L}}\},
\end{equation}%
\newline
are computed up to the first three flows $t_{1},t_{3},t_{5}$. We work out
these equations by adding other flows namely $t_{7}$ and $t_{9}$. We find 
\begin{equation}
\begin{array}{lcl}
{u}_{t_{1}} & = & {u^{\prime }} \\ 
&  &  \\ 
{u}_{t_{3}} & = & \frac{3}{2}uu^{\prime }+{(\frac{1}{2})^{2}}u^{\prime
\prime \prime } \\ 
&  &  \\ 
{u}_{t_{5}} & = & \frac{15}{8}u^{2}u^{\prime }+5{(\frac{1}{2})^{2}}%
(u^{\prime }u^{\prime \prime }+\frac{1}{2}uu^{\prime \prime \prime })+(\frac{%
1}{2})^{4}u^{(5)} \\ 
&  &  \\ 
{u}_{t_{7}} & = & {\frac{35}{16}}u^{3}u^{\prime }+\frac{35}{8}{(\frac{1}{2}%
)^{2}}({4}uu^{\prime }u^{\prime \prime }+u^{\prime 3}+u^{2}u^{\prime \prime
\prime })+\frac{7}{2}(uu^{(5)}+3u^{\prime }u^{(4)}+{5}u^{\prime \prime
}u^{\prime \prime \prime })(\frac{1}{2})^{4}+(\frac{1}{2})^{6}u^{(7)} \\ 
&  &  \\ 
{u}_{t_{9}} & = & 18(\frac{1}{2})^{6}u^{\prime }u^{(6)}+{\frac{651}{8}}(%
\frac{1}{2})^{4}u^{\prime }(u^{\prime \prime })^{2}+{\frac{315}{128}}%
u^{4}u^{\prime }+{\frac{483}{8}}(\frac{1}{2})^{4}u^{\prime }{}^{2}u^{\prime
\prime \prime }+{\frac{315}{16}}(\frac{1}{2})^{2}uu^{\prime }{}^{3}+{\frac{%
189}{4}}{\ }(\frac{1}{2})^{4}uu^{(4)}u^{\prime } \\ 
&  &  \\ 
& + & {\frac{315}{8}}(\frac{1}{2})^{2}u^{2}u^{\prime }u^{\prime \prime }+{%
\frac{315}{4}}(\frac{1}{2})^{4}uu^{\prime }u^{\prime \prime \prime }+63(%
\frac{1}{2})^{6}u^{\prime \prime \prime }u^{(4)}+\frac{105}{16}(\frac{1}{2}%
)^{2}u^{3}u^{\prime \prime \prime }+42(\frac{1}{2})^{6}u^{(5)}u^{\prime
\prime } \\ 
&  &  \\ 
& + & {\frac{63}{8}}(\frac{1}{2})^{4}u^{2}u^{(5)}+(\frac{1}{2})^{8}u^{(9)}+%
\frac{9}{2}(\frac{1}{2})^{6}uu^{(7)}%
\end{array}%
\end{equation}%
\newline
\newline
Ssome important remarks are in order:\newline
\newline
\textbf{1}. The flow parameters $t_{2k+1}$ has the following conformal
dimension $[\partial _{t_{2k+1}}]=-[t_{2k+1}]=2k+1$ for $k=0,1,2,...,$.%
\newline
\newline
\textbf{2}. A remarkable property of the $sl_{2}$ KdV hierarchy is about the
degree of non linearity of the evolution equations Eq.(4.9). We present in
the following table the behavior of the higher non-linear terms with respect
to the first leading flows $t_{1},...,t_{9}$ and give the behavior of the
general flow parameter $t_{2k+1}$.\newline
\newline
\begin{equation}
\begin{tabular}{ccc}
&  &  \\ 
&  &  \\ 
Flows & \hspace{1cm} The higher n.l. terms & \hspace{1cm} Degree of n
linearity \\ 
$t_{1}$ & $u^{0}u^{\prime }=u^{\prime }$ & $0$ \\ 
&  &  \\ 
$t_{3}$ & $\frac{3}{2}uu^{\prime }$ & \hspace{3cm} $1$ \hspace{0cm}
(quadratic) \\ 
&  &  \\ 
$t_{5}$ & $\frac{15}{2^{3}}u^{2}u^{\prime }$ & \hspace{2cm} $2$ \hspace{0cm}
(cubic) \\ 
&  &  \\ 
$t_{7}$ & $\frac{35}{2^{4}}u^{3}u^{\prime }$ & $3$ \\ 
&  &  \\ 
$t_{9}$ & $\frac{315}{2^{7}}u^{4}u^{\prime }$ & $4$ \\ 
&  &  \\ 
... & ... & ... \\ 
&  &  \\ 
$t_{2k+1}$ & \hspace{0cm} $\eta {(2k+1)(2k-1)}u^{k}u^{\prime }$ & \hspace{1cm%
} $(k)$,%
\end{tabular}%
\end{equation}%
where $\eta $ is an arbitrary constant.\newline
\newline
This result shows among others that the evolution equations Eq.(4.9) exhibit
at most a nonlinearity of degree $(k)$ associated to a term proportional to $%
(2k+1)(2k-1)u^{k}u^{\prime }$. The particular case $k=0$ corresponds to
linear wave equation. \newline
\newline
\textbf{3}. The results show a correspondence between the flows $t_{2k+1}$
and the coefficient number $(\frac{1}{2})^{2(k-s)},0\leq s\leq k$.
Particularly, the higher term $(\frac{1}{2})^{2(k)}$ is coupled to the $k-th$
prime derivative of $u_{2}$ namely $u^{(k)}$.\newline
\newline
\textbf{4}. Once the non linear terms in the evolution equations are
ignored, there will be no solitons in the KdV-hierarchy as the latter's are
intimately related to non linearity.

\subsection{$sl_3$-Boussinesq Hierarchy}

The same analysis used in deriving the $sl_{2}$-KdV hierarchy is actually
extended to build the $sl_{3}$-Boussinesq hierarchy. The latter is
associated to the momentum Lax operator ${\mathcal{L}}_{3}=\partial
^{3}+u_{2}\partial +u_{3}$ whose $3-th$ root reads as 
\begin{equation}
\begin{array}{lcl}
{\mathcal{L}}^{\frac{1}{3}} & = & \Sigma _{i=-1}b_{i+1}\partial ^{-i} \\ 
&  &  \\ 
& = & {\Sigma }_{i=-1}a_{i+1}\partial ^{-i}%
\end{array}%
\end{equation}%
in such way that ${\mathcal{L}}_{3}={\mathcal{L}}^{\frac{1}{3}}{\mathcal{L}}%
^{\frac{1}{3}}{\mathcal{L}}^{\frac{1}{3}}$. Explicit computations lead to 
\begin{equation}
\begin{array}{lcl}
b_{0} & = & 1 \\ 
&  &  \\ 
b_{1} & = & 0 \\ 
&  &  \\ 
b_{2} & = & \frac{1}{3}u_{2} \\ 
&  &  \\ 
b_{3} & = & \frac{1}{3}u_{3}-\frac{2}{6}u_{2}^{^{\prime }} \\ 
&  &  \\ 
b_{4} & = & -\frac{1}{9}u_{2}^{2}-\frac{2}{6}u_{3}^{^{\prime }}+\frac{8}{9}(%
\frac{1}{2})^{2}u_{2}^{^{\prime \prime }} \\ 
&  &  \\ 
b_{5} & = & -\frac{2}{9}u_{2}u_{3}+\frac{8}{18}u_{2}u_{2}^{\prime }+\frac{8}{%
9}(\frac{1}{2})^{2}u_{3}^{\prime \prime }-\frac{8}{9}(\frac{1}{2}%
)^{3}u_{3}^{\prime \prime \prime } \\ 
&  &  \\ 
b_{6} & = & \frac{1}{9}\{\frac{5}{9}u_{2}^{3}-u_{3}^{2}+(4u_{2}u_{3}^{\prime
}+5u_{2}^{\prime }u_{3})-20(\frac{1}{2})^{2}(u_{2}u_{2}^{\prime \prime
}+(u_{2}^{\prime })^{2})-8(\frac{1}{2})^{3}u_{3}^{\prime \prime \prime }+%
\frac{16}{3}(\frac{1}{2})^{4}u_{2}^{\prime \prime \prime \prime }\} \\ 
&  &  \\ 
b_{7} & = & \frac{1}{9}\{\frac{5}{3}u_{2}^{2}u_{3}+5(u_{3}u_{3}^{\prime
}-u_{2}^{2}u_{2}^{\prime })-\frac{20}{3}(\frac{1}{2})^{2}(5u_{2}^{\prime
\prime }u_{3}+7u_{2}^{\prime }u_{3}^{\prime }+u_{2}u_{3}^{\prime \prime
\prime }) \\ 
&  &  \\ 
&  & -40(\frac{1}{2})^{3}(3u_{2}^{\prime }u_{2}^{\prime \prime
}+u_{2}u_{2}^{\prime \prime \prime })+\frac{16}{3}(\frac{1}{2}%
)^{4}u_{3}^{\prime \prime \prime \prime }) \\ 
&  &  \\ 
b_{8} & = & {\frac{5}{27}}(u_{{2}}u_{{3}}^{2}-{\frac{2}{9}}u_{{2}}^{4})-%
\frac{5}{9}(u_{2}^{2}u_{3}^{\prime }-\frac{7}{3}u_{2}^{\prime }u_{2}u_{3})+%
\frac{5}{81}(12u_{3}^{2\prime }+31u_{2}u_{2}^{2\prime } \\ 
&  &  \\ 
&  & +17u_{2}^{2}u_{2}^{\prime \prime }-15u_{3}^{\prime \prime }u_{3})+\frac{%
5}{27}(10u_{3}^{\prime \prime }u_{2}^{\prime }+13u_{2}^{\prime \prime
}u_{3}^{\prime }+7u_{3}u_{2}^{\prime \prime \prime }+3u_{3}^{\prime \prime
}u_{2}) \\ 
&  &  \\ 
&  & +\frac{5}{81}(8u_{2}^{4}u_{2}+23u_{2}^{2\prime ^{2}}+32u_{2}^{\prime
}u_{2}^{\prime \prime \prime })+{\frac{1}{81}}u_{{2}}^{(6)} \\ 
&  &  \\ 
&  & 
\end{array}%
\end{equation}%
Similarly, one can easily determine the coefficients $a_{i+1}$ which are
also expressed as functions of $b_{i+1}$ and their derivatives. This result
is summarized in the expression of ${\mathcal{L}}^{\frac{1}{3}}$ namely 
\begin{equation}
\begin{array}{lcl}
{\mathcal{L}}^{\frac{1}{3}}=\partial  & + & \frac{1}{3}u_{2}\partial ^{-1}
\\ 
&  &  \\ 
& + & \frac{1}{3}\{u_{3}-\frac{1}{2}u_{2}^{\prime }\}\partial ^{-2} \\ 
&  &  \\ 
& - & \frac{1}{9}\{u_{2}^{2}+\frac{1}{2}^{2}u_{2}^{\prime \prime }\}\partial
^{-3} \\ 
&  &  \\ 
& + & \frac{1}{9}\{-2u_{2}u_{3}+u_{2}^{\prime }u_{2}-\frac{1}{2}%
^{2}u_{3}^{\prime \prime }+\frac{1}{2}^{3}u_{2}^{\prime \prime \prime
}\}\partial ^{-4} \\ 
&  &  \\ 
& + & \frac{1}{9}\{\frac{1}{3}{\frac{1}{2}}^{4}u_{{2}}^{(4)}+u_{{2}}^{\prime
}u_{{3}}-u_{{3}}^{2}+{\frac{5}{9}}u_{{2}}^{3}\}.{\partial }^{-5} \\ 
&  &  \\ 
& + & \frac{1}{27}\{{5}u_{{2}}^{2}u_{{3}}-{5}\frac{1}{2}u_{{2}}^{2}u_{{2}%
}^{\prime }+\frac{5}{2}(u_{{2}}^{\prime }u_{{3}}^{\prime }-u_{{2}}^{\prime
\prime }u_{{3}})+{\frac{1}{2}}^{4}u_{{3}}^{(4)}-{\frac{1}{2}}^{5}u_{{2}%
}^{(5)}\}.{\partial }^{-6} \\ 
&  &  \\ 
& + & \frac{1}{27}{\big \{}\frac{5}{9}u_{2}(9u_{3}^{2}-2u_{2}^{3})-5u_{{2}%
}^{\prime }u_{{2}}u_{{3}}+\frac{5}{3}\frac{1}{2}^{2}(6u_{3}^{^{\prime
}2}-6u_{3}^{\prime \prime }u_{3}+5u_{2}^{2}u_{2}^{\prime \prime
}-2u_{2}u_{2}^{2\prime }) \\ 
&  &  \\ 
& - & 10{\frac{1}{2}}^{3}(-u_{3}^{\prime \prime }u_{2}^{\prime
}-u_{3}u_{2}^{\prime \prime \prime }+2u_{2}^{\prime \prime }u_{3}^{\prime })-%
\frac{10}{3}{\frac{1}{2}}^{4}(u_{2}^{(4)}u_{2}+4u_{2}^{\prime }u_{2}^{\prime
\prime \prime }-5u_{2}^{\prime \prime ^{2}})-{\frac{1}{3}}{\frac{1}{2}}%
^{6}u_{{2}}^{(6)}{\big \}}{\partial }^{-7} \\ 
&  &  \\ 
& + & ...%
\end{array}%
\end{equation}%
Furthermore, using the $sl_{3}$-Lax evolution equations 
\begin{equation}
\frac{\partial {\mathcal{L}}}{\partial t_{k}}=[{\mathcal{L}}_{+}^{\frac{k}{3}%
},{\mathcal{L}}],
\end{equation}%
that we compute explicitly for $k=1,2,4$ we obtain 
\begin{equation}
\begin{array}{lcl}
\frac{\partial {\mathcal{L}}}{\partial t_{1}} & = & u_{2}^{\prime }\partial
+u_{3}^{\prime }-{\frac{1}{2}}u_{2}^{\prime \prime } \\ 
&  &  \\ 
\frac{\partial {\mathcal{L}}}{\partial t_{2}} & = & 2\{u_{3}^{\prime }-\frac{%
1}{2}u_{2}^{\prime \prime }\}\partial -\frac{2}{3}\{u_{2}u_{2}^{\prime }+(%
\frac{1}{2})^{2}u_{2}^{\prime \prime \prime }\} \\ 
&  &  \\ 
\frac{\partial {\mathcal{L}}}{\partial t_{4}} & = & \frac{4}{3}%
\{(u_{2}u_{3})^{\prime }-\frac{1}{2}(u_{2}^{\prime \prime
}u_{2}+u_{2}^{^{\prime }2})+2(\frac{1}{2})^{2}u_{3}^{\prime \prime \prime
}-2(\frac{1}{2})^{3}u_{2}^{(4)}\}\partial  \\ 
&  &  \\ 
&  & +\frac{4}{3}\{u_{3}u_{3}^{\prime }-\frac{1}{3}u_{2}^{2}u_{2}^{\prime }-%
\frac{1}{2}(u_{2}^{\prime }u_{3}^{\prime }+u_{2}^{\prime \prime }u_{3})-(%
\frac{1}{2})^{2}(u_{2}^{\prime }u_{2}^{\prime \prime }+u_{2}u_{2}^{\prime
\prime \prime })-\frac{2}{3}(\frac{1}{2})^{4}u_{2}^{(5)}\} \\ 
&  &  \\ 
&  & 
\end{array}%
\end{equation}%
Identifying both sides of the previous equations, one obtain the following
first leading evolution equations 
\begin{equation}
\begin{array}{lcl}
\frac{\partial }{\partial t_{1}}{u_{2}} & = & {u_{2}^{\prime }} \\ 
&  &  \\ 
\frac{\partial }{\partial t_{1}}{u_{3}} & = & {u_{3}^{\prime }-\frac{1}{2}}%
u_{2}^{\prime \prime } \\ 
&  &  \\ 
&  &  \\ 
\frac{\partial }{\partial t_{2}}{u_{2}} & = & 2{u_{3}^{\prime }}%
-u_{2}^{\prime \prime } \\ 
&  &  \\ 
\frac{\partial }{\partial t_{2}}{u_{3}} & = & -\frac{2}{3}u_{2}u_{2}^{\prime
}-\frac{2}{3}(\frac{1}{2})^{2}u_{2}^{\prime \prime \prime } \\ 
&  &  \\ 
&  &  \\ 
\frac{\partial }{\partial t_{4}}{u_{2}} & = & \frac{4}{3}\{(u_{2}u_{3})^{%
\prime }-\frac{1}{2}(u_{2}^{\prime \prime }u_{2}+u_{2}^{^{\prime }2})+2(%
\frac{1}{2})^{2}u_{3}^{\prime \prime \prime }-2(\frac{1}{2})^{3}u_{2}^{(4)}\}
\\ 
&  &  \\ 
\frac{\partial }{\partial t_{4}}{(u_{3}-\frac{1}{2}u_{2}^{\prime })} & = & 
\frac{4}{3}\{u_{3}u_{3}^{\prime }-\frac{1}{3}u_{2}^{2}u_{2}^{\prime }-\frac{1%
}{2}(u_{2}^{\prime }u_{3}^{\prime }+u_{2}^{\prime \prime }u_{3})-(\frac{1}{2}%
)^{2}(u_{2}^{\prime }u_{2}^{\prime \prime }+u_{2}u_{2}^{\prime \prime \prime
})-\frac{2}{3}(\frac{1}{2})^{4}u_{2}^{(5)}.%
\end{array}%
\end{equation}%
These equations define what we call the $sl_{3}$-Boussinesq hierarchy. The
first two equations are simply linear independent wave equations fixing the
dimension of the first flow parameter $t_{1}$ to be $[t_{1}]=-1$. \newline
\newline
The non trivial flow of this hierarchy starts really from the second couple
of equations associated to $t_{2}$. It's important to point out that its
important to deal with the basis of primary conformal fields $v_{k}$ instead
of the old basis $u_{k}$ \cite{DIZ, ss94}, one can write the previous couple
of equations in term of the spin $3$ primary field $v_{3}=u_{3}-\frac{1}{2}%
u_{2}^{\prime }$ as follows 
\begin{equation}
\begin{array}{lcl}
\frac{\partial }{\partial t_{2}}{u_{2}} & = & 2{v_{3}^{\prime }} \\ 
&  &  \\ 
\frac{\partial }{\partial t_{2}}{v_{3}} & = & -\frac{2}{3}%
\{u_{2}u_{2}^{\prime }+(\frac{1}{2})^{2}u_{2}^{\prime \prime \prime }\}%
\end{array}%
\end{equation}%
This couple of equations define the Boussinessq equation. Its second-order
form is obtained by differentiating the first equation in Eq.(4.17) with
respect to $t_{2}$ and then using the second equation. We find 
\begin{equation}
\frac{\partial ^{2}}{\partial t_{2}^{2}}u_{2}=-\frac{4}{3}%
(u_{2}u_{2}^{\prime }+(\frac{1}{2})^{2}u_{2}^{(3)})^{\prime },
\end{equation}

Recall that the classical Boussinesq equation is associated to the $sl_{3}$%
-Lax differential operator 
\begin{equation}
{\mathcal{L}}_{3}=\partial ^{3}+2u\partial +v_{3}
\end{equation}%
with $v_{3}=u_{3}-\frac{1}{2}u_{2}^{\prime }$ defining the spin-$3$ primary
field. This equation which takes the following form 
\begin{equation}
u_{tt}=-(auu^{\prime }+bu^{(3)})^{\prime },
\end{equation}%
where $a,b$ are arbitrary constants, arises in several physical
applications. Initially, it was derived to describe propagation of long
waves in shallow water. This equation plays also a central role in $2d$
conformal field theories via its Gelfand-Dickey second Hamiltonian structure
associated to the Zamolodchikov $w_{3}$ non linear algebra.\newline
\newline
Similarly the third couple of equations Eq.(4.16) can be equivalently
written as 
\begin{equation}
\begin{array}{lcl}
\frac{\partial }{\partial t_{4}}{u_{2}} & = & \frac{4}{3}(u_{2}v_{3}+2(\frac{%
1}{2})^{2}v_{3}^{\prime \prime })^{\prime } \\ 
&  &  \\ 
\frac{\partial }{\partial t_{4}}{v_{3}} & = & \frac{4}{3}\{v_{3}v_{3}^{%
\prime }-(\frac{1}{2})^{2}u_{2}u_{2}^{\prime \prime \prime }-\frac{1}{3}%
(u_{2}^{2}u_{2}^{\prime }+2(\frac{1}{2})^{4}u_{2}^{(5)})\}%
\end{array}%
\end{equation}%
To close work note that other flows equations associated to $(sl_{2})$-KdV
and $(sl_{3})$-Boussinesq hierarchies can be also derived once some lengthly
and hard computations are performed. One can also generalize the obtained
results by considering other $sl_{n}$ integrable hierarchies with $n>3$.


\begin{thebibliography}{99}
\bibitem{1} L.D.Faddeev, L.A.Takhtajan, Hamiltonian Methods and the theory
of solitons, 1987, \newline
E. Date, M. Kashiwara, M. Jimbo and T. Miwa in "Nonlinear Integrable
Systems", eds. M. Jimbo and T. Miwa, World Scientific (1983), \newline
A. Das, Integrable Models (World Scientific, Singapore, 1989).

\bibitem{cft} A.A.Belavin, A.M.Polyakov, A.B.Zamolodchikov,
Nucl.Phys.B241(1984);\newline
V. S. Dotsenko, V.A. Fateev, Nucl Phys. B240 [FS12](1984)312;\newline
C.~Itzykson, H.~Saleur and J.~B.~Zuber, 
Europhys.\ Lett.\ \textbf{2}, 91 (1986); 
\newline
P. Ginsparg, Applied Conformal field Theory, Les houches Lectures (1988).

\bibitem{string} Becker, K., M. Becker, and J. Schwarz, String Theory and
M-Theory: A Modern Introduction, Cambridge University Press, New York, 2007,%
\newline
Green, M., J. Schwarz, and E. Witten, Superstring Theory, vol. 1:
Introduction (Cambridge Monographs on Mathematical Physics), Cambridge
University Press, New York, 1988.

\bibitem{bak} I. Bakas, Nuc. Phys. B302, 189(1988).

\bibitem{GN} {\small 
J.~L.~Gervais, 
Phys.\ Lett.\ B \textbf{160}, 277 (1985). 
}\newline
{\small 
A.~Bilal and J.~L.~Gervais, 
Phys.\ Lett.\ B \textbf{206}, 412 (1988). 
}

\bibitem{dubrovin} B. Dubrovin, "Hamiltonian perturbations of hyperbolic
PDEs and applications, Workshop on Integrbale systems and scientifc
computing, 15-20 June 2009",

\bibitem{w1} A. B Zamolodchikov, Teor. Math. Fiz 65, 347 (1985),\newline
V. A. Fateev and A. B. Zamolodchikov, Nucl. Phys. B 304, 348 (1988).

\bibitem{w2} K. Schoutens, A. Servin, and P. Van Nieuvenhuizen, Phys. Len. B
243, 245 (1990),\newline
E. Bergshoeff, A. Bilal, and K. S. Stelle, TH 5924/90,\newline
C. M. Hull, Talk given at the summer workshop on high energy physics, 1992.

\bibitem{w3} V. A. Fateev and S. Lukyanov, Int. J. Mod. Phys. A 3, 5O7
(1987),\newline
E Bais, P. Bouwknegt, M. Surridge, and K. Schoutens, Nucl. Phys.B 304, 348
(1988),\newline
L. Romans, Nucl. Phys.B 352, 829 (1991),\newline
E.Bergshoeff, C.N. Pope, L.J.Romans, E.Sezgin, X.Shen, Phys. Lett. B 245,
1990.

\bibitem{DIZ} P. Di Francesco, C. Itzykson, and J.-B. Zuber, Commun. Math.
Phys. 148, 543 (1991).

\bibitem{SE1} F. Magri, J. Math. Phys. 19(1978) 1156. 

\bibitem{SE2} J.~C.~Brunelli and A.~K.~Das, 
arXiv:hep-th/9410165. 

\bibitem{TB} B.A. Kupershmidt, Commun. Math. Phys. 99 (1985) 51.

\bibitem{ss94} E.H. Saidi and M.B. Sedra, J. Math. Phys. 35(1994)3190

\bibitem{tod1} Yu.I.Manin and A.O.Radul, Cummun. Math. Phys. 98(1985)65,%
\newline
K.Yamagishi, Phys.Lett.B.205 (1988)466,\newline
P.Mathieu, Phys.Lett.B.208,101 (1988); Jour.Math. Phys 29 (1988) 2499,%
\newline
I.Bakas, Phys Lett.B. 219 (1989) 283; Comm.Math .Phys. 123 (1989) 627,%
\newline
J.D.Smit, Comm. Math. Phys. 128 (1990)1

\bibitem{tod2} P. Mansfield, Nucl. Phys.B 208 (1982)277; B222(1983)419%
\newline
D.Olive and N.Turok, Nucl. Phys. B 257[FS14](1986)277.

\bibitem{tod3} E.H. Saidi and M.B. Sedra, Int. Jour. Mod. Phys.
A9(1994)891-913;
\end{thebibliography}
\end{document}